\documentclass[prd,aps,amsfonts,eqsecnum,nofootinbib,longbibliography,notitlepage]{revtex4-1}

\usepackage{soul}
\usepackage{graphicx}
\usepackage{xcolor}
\usepackage{caption}
\usepackage{rotating}
\usepackage{amsmath,amssymb,graphics,amsthm,isomath}
\usepackage{subfigure}
\usepackage{physics }

\usepackage[colorlinks=true, urlcolor=violet, linkcolor=blue, citecolor=red, hyperindex=true, linktocpage=true]{hyperref}
\usepackage[capitalise,compress]{cleveref}

\numberwithin{thm}{section}

\makeatletter
\renewcommand{\p@subsection}{}
\renewcommand{\p@subsubsection}{}
\makeatother

\usepackage{xcolor}
\usepackage{mathtools}

\usepackage{dsfont}

\begin{document}
\setstcolor{red}

\title{Locality bounds for quantum dynamics at low energy}

\author{Andrew Osborne}
\email{andrew.osborne-1@colorado.edu}
\affiliation{Department of Physics and Center for Theory of Quantum Matter, University of Colorado, Boulder CO 80309, USA}

\author{Chao Yin}
\email{chao.yin@colorado.edu}
\affiliation{Department of Physics and Center for Theory of Quantum Matter, University of Colorado, Boulder CO 80309, USA}

\author{Andrew Lucas}
\email{andrew.j.lucas@colorado.edu}
\affiliation{Department of Physics and Center for Theory of Quantum Matter, University of Colorado, Boulder CO 80309, USA}

\begin{abstract}
    We discuss the generic slowing down of quantum dynamics in low energy density states of spatially local Hamiltonians.  Beginning with quantum walks of a single particle, we prove that for certain classes of Hamiltonians (deformations of lattice-regularized $H\propto p^{2k}$), the ``butterfly velocity" of particle motion at low energies  has an upper bound that must scale as $E^{(2k-1)/2k}$, as expected from dimensional analysis.  We generalize these results to obtain bounds on the typical velocities of particles in many-body systems with repulsive interactions, where for certain families of Hubbard-like models we obtain similar scaling.  
\end{abstract}

\date{\today}
\maketitle

\section{Introduction}

In classical statistical mechanics, the slowdown of the dynamics of particles at any finite temperature is well--understood. 
The Maxwell--Boltzmann distribution tells us that the probability that any one particle moves very fast gets larger as temperature increases. 
As an explicit example, consider \begin{equation}
    H = \sum_{i=1}^N \frac{p_i^2}{2m} + \sum_{i<j} V(x_i-x_j),
\end{equation}
with some repulsive interactions $V\ge 0$. 
Let us ask about the behavior of this system in a state drawn from the Gibbs ensemble, where the probability density of seeing the system in configuration $(x_i,p_i)$ is given by $\rho = Z^{-1}\mathrm{e}^{-\beta H(x,p)}$, where $Z$ (the partition function) is an overall normalization constant.
If $V=0$, the typical momentum squared $\langle p^2\rangle \sim T$ $v_{\text{rms}} \sim \sqrt{T}$.  Even in the presence of $V>0$ however, the dynamics necessarily slows down at sufficiently low temperature: the probability measure on phase space becomes more and more concentrated at low $T$, and on the momenta phase space coordinates $p_i$, we are more and more likely to be found close to $p_i=0$.  This is despite the fact that, \emph{in principle}, with energy conserving dynamics, a finite fraction of the total energy could be pushed into the kinetic energy of particle 1.  Such a configuration is allowed in phase space, yet the probability of finding this is exponentially small with $N$.  In quantum mechanics, we could similarly argue that the probability at any one fixed time, the probability for one particle to have very large kinetic energy is small.


  However, it is more challenging to prove that information cannot propagate with faster velocities, because information may spread through correlations, which in turn are ``propagated" due to the rare fast particles in the system. It has been a mathematical challenge for a long time to confirm the classical intuition that physics in low energy states is ``slower" than high energy states in typical systems.  We intuitively expect this to be true for most systems, except those with (emergent) Lorentz invariance, and those with a Fermi surface.  Yet as far as we can tell, the only known rigorous results along these lines, capable of proving that velocities \emph{vanish} as energy vanishes, are for \emph{single particle} problems, perhaps obeying nonlinear Schr\"odinger equations \cite{Jensen1985PropagationEF,skibsted,huang2019uncertainty,Arbunich_2021,rev1,rev2,rev3}. And even for a single particle, most results are in the context of the textbook nonrelativistic Schr\"odinger equation, and do not consider dynamics on lattices or with more non-trivial kinetic energies.

  One mathematically rigorous result on the slowdown of dynamics at low energies is obtained in \cite{Han_2019}; however, their results do not in general guarantee that velocities must vanish at low temperature, even in models where on physical grounds this is anticipated.  Indeed, there are non-rigorous calculations \cite{blake_holo,swingle}, using gauge-gravity duality and/or quantum field theory, which show that in a \emph{many-particle} theory with dynamical critical exponent $z$ (i.e. the relative scalings of time $t$ and space $x$ is $t\sim x^z$), the emergent velocity scale for low temperature dynamics is \begin{equation}
      v_{\mathrm{B}} \sim T^{1-1/z}. 
  \end{equation}
  Here $v_{\mathrm{B}}$ is called the butterfly velocity, and is usually measured using out-of-time-ordered correlation functions.  From our perspective, these correlators intuitively bound the growth of operators and the spreading of quantum information within the low energy subspace. (\ref{eq:vB}) can be intuitively understood as follows. Consider a single-particle theory with energy-momentum dispersion relation:
\begin{equation}
    E \sim p^z. \label{eq:Epz}
\end{equation}
A particle in a system at temperature $T$ has energy $E \sim T$, but it spreads out on a scale of length $\xi \sim p^{-1}$. By dimensional analysis, if $p^z \sim T$, then $p \sim T^{\frac{1}{z}}$, so we expect 
\begin{equation}\label{eqn:scaling}
    v_{\mathrm B} \lesssim T \xi \sim T^{1 - \frac{1}{z}}.
\end{equation}
This intuition and conjecture were originally provided by \cite{blake_holo,swingle}. 
Yet another intuitive argument is that information should travel at the group velocity
\begin{equation}
    v_{\mathrm{g}} \sim p^{z - 1} \sim T^{1 - \frac{1}{z}}.
\end{equation}

Most rigorous results about the spreading of quantum information in many-body systems  take the form of Lieb--Robinson bounds \cite{Lieb1972TheFG,Hastings_2006,Nachtergaele_2006}. These bounds apply to \emph{all} quantum states, and therefore cannot capture the slowdown of dynamics at low energies: see \cite{chen2023speed} for a recent review.  
Despite much progress in generalizing Lieb-Robinson bounds in recent years, few of the techniques explain how dynamics can slow down in low energy or finite density states. 

In this paper, we study the dynamics of a single particle in the \emph{low energy} subspace of single-particle -- and certain multi-particle -- problems.  This simplified setting is meant to capture the essence of the low temperature Gibbs ensemble, but it is \emph{not equivalent} to a mathematical proof of slowdown at low temperature. Still, in this drastically simplified setting, we prove that the (typical) particle cannot move faster than an effective velocity which scales as \begin{equation}
      v_{\mathrm{B}} \sim E^{1-1/z}. \label{eq:vB}
  \end{equation}
  where $E$ denotes the energy in the single-particle problem, or the average energy per particle in the restricted class of many-body problems that we study.  Intuitively $E$ plays the same role as temperature $T$ in the conjectured bounds described previously. Our main goal is to emphasize that at least in certain settings, the techniques developed originally in the mathematics literature to study single-particle dynamics \cite{Jensen1985PropagationEF,skibsted,huang2019uncertainty,Arbunich_2021} may be relevant for many-body problems \cite{rev1} in low-energy ensembles.   Recent work has also made this connection in the specialized setting of number-conserving bosonic dynamics \cite{Faupin:2021nhk,Faupin:2021qpp}; see \cite{rahul2020,lowDensity2020,lucas_finite_density} for an alternative perspective on slow physics in number-conserving models at low density.

In what follows, we define a generalization of the translation--invariant models described above -- both in the continuum, and in lattice regularizations -- where there is a reasonable definition of $z$, and where we demonstrate that such models obey (\ref{eq:vB}).  Most of the paper will focus on the single-particle problem, since, as we will show, these bounds allow us to immediately constrain certain many-body problems at the end of the paper.

\section{One particle in the continuum}\label{sec:continuum}
We begin by studying the dynamics of a single particle in the continuum.   The study of Lieb-Robinson-like bounds is well-established here for the ordinary Schr\"odinger equation: see e.g. \cite{Jensen1985PropagationEF,skibsted,huang2019uncertainty,Arbunich_2021}. This section contains a mild generalization of these results to a more general family of continuum Hamiltonians.
\subsection{One dimension}\label{sec:oned}
We start with one spatial dimension.  Take the operators $p$ and $x$ to be canonical conjugates: $[x,p]=\mathrm{i}$, and 
\begin{equation}
    p = - \mathrm i \frac{\partial}{\partial x}.
\end{equation}
We consider the Hermitian Hamiltonian operator
\begin{equation}\label{eqn:contham}
  H = p^\alpha f(x) p^\alpha + V(x),
\end{equation}
with the requirements that
\begin{equation}\label{eqn:fband}
 0 < a \leq f(x) \leq b < \infty
\end{equation}
and 
\begin{equation}
  V(x) \geq 0.
\end{equation}
Further, we demand that $\alpha$ be a positive integer. This assumption will be made whenever hereafter $\alpha$ appears.
Such models are a reasonable generalization of translation invariant models with critical dynamical exponent $z = 2 \alpha$.  After all, if $f$ and $V$ are constant functions, $H$ has plane wave eigenfunctions $\mathrm{e}^{\mathrm{i}kx}$ of eigenvalue $f \times k^{2\alpha}$, which is precisely the dispersion relation (\ref{eq:Epz}).

Models of the form (\ref{eqn:contham}) are a coarse--grained analogue of ``random--bond models" \cite{rbond}, which have hopping terms which fluctuate from edge to edge.  If these fluctuations are strongly correlated on short distances, then we may expect that the continuum limit of such a model takes the form of (\ref{eqn:contham}). Later, we will also consider lattice models of a similar form that are more closely analogous to such random--bond models.  

Let the exact (possibly non-normalizable) eigenstates of (\ref{eqn:contham}) be denoted as $|E \rangle$ so that 
\begin{equation}
    H |E\rangle = E |E\rangle .
\end{equation}
Choose some $\epsilon > 0$ and define the projection onto low-energy states:
\begin{equation}\label{eqn:microcanonical}
    \mathbb P_\epsilon |E\rangle  = \mathbb{I}(E\le \epsilon) |E\rangle .
\end{equation}
$\mathbb{P}_\epsilon$ given in (\ref{eqn:microcanonical}) is a kind of hybrid between a canonical and microcanonical (unnormalized) density matrix.
In what follows, we will rigorously bound correlation functions which are projected into the low-energy subspace by $\mathbb P_\epsilon$.   

Since $V(x) \geq 0$ by assumption, we find that 
 \begin{equation}
   \langle\varphi| \mathbb P_\epsilon  p^\alpha f(x) p^\alpha \mathbb P_\epsilon | \varphi\rangle \leq \epsilon
\end{equation}
if \begin{equation}
    |\varphi\rangle = \mathbb{P}_\epsilon |\varphi\rangle. \label{eq:Pepsilonprojector}
\end{equation}
Further, because we have assumed $f(x)$ to take on values only in some positive, finite range, we have that 
\begin{equation}\label{eqn:pbound}
  a \langle\varphi|\mathbb P_\epsilon p^{2 \alpha} \mathbb P_\epsilon|\varphi\rangle \leq \langle\varphi| \mathbb P_\epsilon  p^\alpha f(x) p^\alpha \mathbb P_\epsilon | \varphi\rangle.
\end{equation}
Jensen's inequality says that, for a positive semidefinite operator $M$ and any $k \geq 1$,
 \begin{equation}
   \langle M \rangle^k  \leq \langle M^k \rangle.
\end{equation}
Choosing
  $M = p^{2(\alpha - n)}$
and 
  $k = \frac{\alpha}{\alpha - n}$,
it follows that if $\langle \varphi|\varphi\rangle = 1$,
\begin{equation}
  \langle \varphi| p^{2(\alpha - n)} |\varphi \rangle \leq \langle \varphi | p^{2 \alpha} | \varphi \rangle^{1 - \frac{n}{\alpha}} \leq \left(\frac{\epsilon}{a}\right)^{1 - \frac{n}{\alpha }}
\end{equation}
for any integer $n$ that does not exceed $\alpha$.
By the idempotence of $\mathbb P_\epsilon$,
\begin{equation}\label{eqn:powerbound}
  |\langle \psi| \mathbb P_\epsilon p^{2(\alpha - n)} \mathbb P_\epsilon |\psi \rangle| \leq \left(\frac{\epsilon}{a}\right)^{1 - \frac{n}{ \alpha}} \langle \psi | \mathbb P_\epsilon |\psi \rangle.
\end{equation}
for any state $|\psi\rangle $.

Now we use these inequalities to constrain the spreading of information.  As a simple application of Ehrenfest's theorem, we may write 
\begin{equation}\label{eqn:erh}
  \frac{\mathrm d}{\mathrm d t}\langle x \rangle = -\mathrm i\langle [x,H] \rangle 
\end{equation}
where the $\langle \cdot \rangle$ is taken with a particular but arbitrarily chosen state.  
If we define
\begin{equation}
x(t) = \mathrm{e}^{ \mathrm{i} H t} x \mathrm{e}^{-\mathrm i H t},
\end{equation}
and 
\begin{equation}
    v = \frac{\mathrm d x}{\mathrm d t},
\end{equation}
our goal is to bound $\langle v \rangle$ where the expectation value is taken with respect to a state obeying (\ref{eq:Pepsilonprojector}).
With the Hamiltonian given in (\ref{eqn:contham}), 
\begin{equation}\label{eqn:contcom}
  v = -\mathrm i [x,H] = \alpha p^{\alpha -1 } \{f(x),p\} p^{\alpha - 1}.
\end{equation}
We now wish to bound $\langle \varphi | v | \varphi\rangle$.  Notice that
\begin{equation}
  |\langle \varphi | \mathbb P_\epsilon  p^\alpha f(x) p^{\alpha - 1 } \mathbb P_\epsilon| \varphi \rangle | \leq 
  \left|\langle \varphi | \mathbb P _\epsilon p^{ \alpha} f(x) p^\alpha \mathbb P_\epsilon |\varphi \rangle \langle \varphi |\mathbb P_\epsilon p^{\alpha -1 } f(x)p^{\alpha -1}\mathbb P_\epsilon |\varphi \rangle \right|^{\frac{1}{2}}
\end{equation}
by Cauchy-Schwarz.
Further,
\begin{equation}\label{eqn:cauchy_1}
  |\langle \varphi | \mathbb P_\epsilon p^{\alpha} f(x) p^\alpha \mathbb P_\epsilon |\varphi \rangle |\leq   \epsilon |\langle \varphi| \mathbb P_\epsilon |\varphi \rangle |
\end{equation}
and
\begin{equation}\label{eqn:cauchy_2}
  |\langle \varphi | \mathbb P_\epsilon p^{\alpha -1 } f(x) p^{\alpha - 1}\mathbb P_\epsilon |\varphi \rangle| \leq b |\langle \varphi |\mathbb P_\epsilon p^{2 \alpha - 2} \mathbb P_\epsilon |\varphi \rangle |\leq b\left(\frac{1}{a} \epsilon\right)^{1 - \frac{1}{\alpha}} \langle \varphi |\mathbb P_\epsilon |\varphi \rangle.
\end{equation}
by (\ref{eqn:fband}) and (\ref{eqn:powerbound}). 
Combining (\ref{eqn:cauchy_1}) and (\ref{eqn:cauchy_2}),  
\begin{equation}
  |\langle \varphi | \mathbb P_\epsilon p^\alpha f(x) p^{\alpha - 1} \mathbb P_\epsilon |\varphi \rangle | \leq 
  \sqrt {b a} \left(\frac{1}{a} \epsilon\right)^{1 - \frac{1}{2\alpha}} |\langle \varphi |\mathbb P_\epsilon |\varphi \rangle|. 
\end{equation}
Therefore,
\begin{equation}\label{eqn:contresult}
  \left| \langle \varphi | \mathbb P_\epsilon v(t) \mathbb P_\epsilon|\varphi \rangle \right| \leq 2\alpha  \sqrt{b} a^{\frac{1 - \alpha}{2\alpha}}   \epsilon^{1 -\frac{1}{2\alpha}} \langle \varphi | \mathbb P_\epsilon |\varphi \rangle 
\end{equation}
for any choice of $|\varphi\rangle$.  Since $\epsilon \sim T$, this reproduces the scaling of (\ref{eq:vB}).

The careful reader may notice the factor of $a^{\frac{1-\alpha}{2\alpha}}$ in the bound on $v_\mathrm B$ when $\alpha > 1$. This factor is necessary due to the following effect.  If $a$ is small, there is some spatial region where $f(x)$ is similarly small, and in said region, particles incur a small kinetic energy, while having, in principle, a very large value of $\langle p^{2\alpha}\rangle$. 
In particular, if there is a region where $f(x)$ vanishes, particles can move arbitrarily quickly in said region while still having a small energy, as measured by (\ref{eqn:contham}). It is for this reason that (\ref{eqn:contresult}) diverges when $a \rightarrow 0$.

Using our velocity bounds, it is easy to use Markov's inequality to show that the time it takes for a quantum wave packet propagate a distance $L$ scales as $L/v$.  What is more non-trivial is to get sharp tail bounds on what fraction of the wave function can lie outside of the apparent ``light cone", and much of the work on studying the propagation of single-particles for the ordinary Schr\"odinger equation focuses on such more sophisticated bounds for models with $f=1$ \cite{Jensen1985PropagationEF,skibsted,huang2019uncertainty,Arbunich_2021}.  We were not able to use the techniques derived above to obtain strong tail bounds on the evolution of wave packets obeying (\ref{eq:Pepsilonprojector}).

\subsection{$d>1$ spatial dimensions}\label{sec:nd}
Now, we generalize to the case of continuum models in $d$ spatial dimensions. Let $[x_i, p_j] = \mathrm{i} \delta_{ij}$; we write $p$ with no subscript  to mean total momentum, so that $p^2 = \sum_{i=1}^d p_i^2$.
We will study models of the form 
\begin{equation}
  H = p^\alpha f(x_1,x_2,\dots,x_d) p^\alpha  + V(x_1,x_2,\dots,x_d).
\end{equation}
Such models are only spatially local if $\alpha$ is an even integer, and so we first focus on this case; models with ``odd" $\alpha$ will be described subsequently. We require again (\ref{eqn:fband}).
We will suppress the explicit dependence of $f$ and $V$ on $x_i$ in what follows.
From the argument in Section \ref{sec:oned}, we see 
\begin{equation}\label{eqn:djen}
  \langle \varphi | \mathbb P_\epsilon p^{2(\alpha-n)} \mathbb P_\epsilon |\varphi \rangle \leq  \left(\frac{\epsilon}{a}\right)^{1 - \frac{n}{\alpha}}\langle \varphi | \mathbb P_\epsilon |\varphi \rangle
\end{equation}
because of Jensen's inequality.  
It follows from (\ref{eqn:djen}) that 
\begin{equation}\label{eqn:inhom}
  \langle \varphi | \mathbb P_\epsilon p^{2(\alpha - n) - 2}p_i^2  \mathbb P_\epsilon |\varphi \rangle \leq \left(\frac{\epsilon}{a}\right)^{1 - \frac{n}{\alpha}}\langle \varphi |\mathbb P_\epsilon|\varphi \rangle
\end{equation}
since $p^2 - p_i^2 $ is positive semidefinite. 
From direct calculation,
\begin{equation}
  v_i =- \mathrm i [x_i ,H ] =  \alpha p^{\alpha - 2} (p_i f p^2 + p^2 f^2 p_i )  p^{\alpha - 2 }.
\end{equation}
Proceeding as before,
\begin{equation}\label{eqn:firstnd}
  |\langle \varphi | \mathbb P_\epsilon p^{\alpha - 2}  p_i f p^\alpha \mathbb P_\epsilon |\varphi \rangle| \leq \left(
    b\langle \varphi | \mathbb P_\epsilon p^{2(\alpha - 2)}p_i^2 \mathbb P_\epsilon |\varphi \rangle 
    \langle \varphi |\mathbb P_\epsilon p^\alpha f p^\alpha \mathbb P_\epsilon |\varphi \rangle
  \right)^{\frac{1}{2}}
\end{equation}
by Cauchy--Schwarz. 
First using the triangle inequality and (\ref{eqn:firstnd}), and then squaring, we acquire 
\begin{equation}\label{funnytrick}
    |\langle \varphi | \mathbb P_\epsilon v_i \mathbb P_\epsilon|\varphi\rangle|^2 \leq 4 \alpha^2
    b\langle \varphi |\mathbb P_\epsilon p^\alpha f p^\alpha \mathbb P_\epsilon |\varphi \rangle \times 
     \langle \varphi | \mathbb P_\epsilon p^{2(\alpha - 2)}p_i^2 \mathbb P_\epsilon |\varphi \rangle  
\end{equation}
Summing (\ref{funnytrick}) and using (\ref{eqn:djen}), we immediately find 
\begin{equation}\label{eqn:ddresult}
  \left(\sum_{i=1}^d \left| \langle \varphi | \mathbb P _ \epsilon v_i(t) \mathbb P_\epsilon |\varphi \rangle \right |^2 \right)^{\frac{1}{2}} \leq 2 \alpha \sqrt{ b}\, a^{\frac{1 - \alpha}{2\alpha}} \epsilon^{1 - \frac{1}{2\alpha}}\langle \varphi | \mathbb P_\epsilon |\varphi \rangle .
\end{equation}

In the preceeding paragraph, we considered models only with $\alpha$ even. It should be no suprise that an analogous result would hold with $\alpha$ odd, and we repeat the demonstration with the appropriate redefinitions here.  For simplicity, we again consider an isotropic model, although our technique could generalize to anisotropic models as well. 
Suppose 
\begin{equation}
H  = \sum_{j=1}^d p^{\alpha - 1} p_j f p_j p^{\alpha - 1} + V
\end{equation}
with $\alpha$ odd. For brevity, write 
\begin{equation}
\tilde{f} = \sum_{j=1}^d p_j f p_j.
\end{equation}
For a particular    $i$,
\begin{equation} \label{eqn:ddjen}
    \langle \varphi | \mathbb P_\epsilon p^{\alpha - 3 } p_i \tilde{f} p_i p^{\alpha - 3} \mathbb P_\epsilon |\varphi \rangle  \leq b  \sum_{j = 1}^d \langle \varphi | \mathbb P _\epsilon p^{\alpha - 3} p_i^2 p_j^2 p^{\alpha - 3} \mathbb P_\epsilon |\varphi \rangle  = b\langle \varphi | \mathbb P_\epsilon p^{2 \alpha - 4} p_i^2 \mathbb P_\epsilon | \varphi \rangle
\end{equation}
since every term is positive semidefinite. In order to produce a workable bound, we may then apply (\ref{eqn:inhom}) directly. 
It is now necessary to compute the time derivative of $x_i$, and it is no greater challenge to evaluate the commutator than it was before:
\begin{equation}\label{eqn:ddcom}
   v_i(t) = (\alpha - 1)p^{\alpha - 3} p_i \tilde f p^{\alpha - 1} + (\alpha-1) p^{\alpha -1} \tilde f p_i p^{\alpha-3} + p^{\alpha - 1} \{f,p_i\} p^{\alpha -1}.
\end{equation}
We proceed by bounding every term in (\ref{eqn:ddcom}) individually by using Cauchy--Schwarz and Jensen's inequality in turn. 
Using (\ref{eqn:inhom}) and (\ref{eqn:ddjen}), we find that the first and second term in (\ref{eqn:ddcom}) give
\begin{equation}\label{eqn:normal}
    \left|\langle  \varphi | \mathbb P_\epsilon p^{\alpha-3}p_i \tilde f p^{\alpha-1} \mathbb P _\epsilon|\varphi \rangle \right | \leq \left(
    \left| \langle \varphi | \mathbb P_\epsilon p^{\alpha-3} p_i \tilde f p_i p^{\alpha-3}\mathbb P_\epsilon |\varphi \rangle \right| 
    \left| \langle \varphi | \mathbb P_\epsilon p^{\alpha-1} \tilde f p^{\alpha-1} \mathbb P_\epsilon |\varphi \rangle \right|
    \right)^{\frac{1}{2}} \leq \sqrt{b} a^{\frac{1 - \alpha}{2 \alpha}} \epsilon^{1 - \frac{1}{2\alpha} } \langle \varphi |\mathbb P_\epsilon | \varphi \rangle
\end{equation}
while the last term in (\ref{eqn:ddcom}) gives
\begin{equation}\label{eqn:wierd}
 \left| \langle \varphi | \mathbb P _\epsilon p^{\alpha-1} p_i f p^{\alpha-1} \mathbb P_\epsilon | \varphi \rangle \right | \leq 
 b \left( \left| \langle \varphi |\mathbb P_\epsilon p^{2 \alpha - 2} \mathbb P_\epsilon |\varphi \rangle \right| \left| \langle \varphi | \mathbb P_\epsilon p^{2\alpha - 2} p_i^2 \mathbb P _\epsilon |\varphi \rangle\right|\right)^{\frac{1}{2}} \leq b \left(\frac{\epsilon}{a}\right)^{1 - \frac{1}{2\alpha}} \langle \varphi |\mathbb P_\epsilon |\varphi \rangle.
\end{equation}
Combining (\ref{eqn:normal}) and (\ref{eqn:wierd}) with (\ref{eqn:ddcom}), we ascertain 
\begin{equation}
 \left| \langle\varphi| \mathbb P _\epsilon v_i(t)\mathbb P _\epsilon | \varphi \rangle \right |  \leq \left( 2(\alpha-1)\sqrt{b} a^{\frac{1-\alpha}{2\alpha}} + 2 b a^{\frac{1 - 2\alpha}{2\alpha}}\right)  \epsilon^{1 - \frac{1}{2\alpha}}\langle \varphi | \mathbb P _\epsilon|\varphi\rangle .
\end{equation}
Though this result is superficially different from (\ref{eqn:ddresult}), it shares the same scaling in $a$ and $\epsilon$.

\section{Lattice Calculation}\label{sec:lattice}

We turn our attention to lattice models. For simplicity, we focus on one dimensional models here, but a generalization to higher dimensions is possible. The first issue to address is to define a momentum analogue to $p$, in order to generalize $p^\alpha f p^\alpha$ to the lattice.
Define the discrete derivative
\begin{equation}
  \nabla =- \mathrm i\sum_{n = -\infty}^\infty \left[- |n \rangle \langle n | +  |n \rangle \langle n+1 |\right]
\end{equation}
and 
\begin{equation}
  x = \sum_{n = -\infty}^\infty n |n \rangle \langle n | .
\end{equation}
We take $x$ and $\nabla$ to be positon and momentum analogues.
Notice that 
\begin{equation}\label{eqn:comm}
  [x, \nabla] =\mathrm i \mathbb I -\nabla
\end{equation}
with $\mathbb{I}$ the identity matrix.  We regard (\ref{eqn:comm}) as a natural analogue of the canonical commutation relations; i.e. 
this is the closest that we can find to a useful notion of $x$ and $p$.  On a countable Hilbert space, it is impossible to write a matrix 
 \begin{equation}
   \Gamma = \sum_{n,m} c_{nm} |n \rangle \langle m |
\end{equation}
so that 
\begin{equation}
  [\Gamma, x] = \mathrm i \mathbb I
\end{equation}
since the diagonal elements of $[\Gamma,x] $ must vanish. 

Hence, we write 
\begin{equation}\label{eqn:dham}
  H = \nabla^{\dagger\alpha} f(x) \nabla^\alpha + V(x)
\end{equation}
where again we require that  $f$ is bounded below by $a$ and bounded above by $b$.
and we also require that $V$ commutes with $x$ and is positive semidefinite. 
We retain the previous definition of $\mathbb P_\epsilon $ with (\ref{eqn:dham}) in place of (\ref{eqn:contham}), and focus on correlators in states where  (\ref{eq:Pepsilonprojector}) holds.
Since 
\begin{equation}\label{eqn:interm}
  a \langle \varphi | \nabla^{\dagger\alpha}  \nabla^\alpha |\varphi \rangle \leq \epsilon
\end{equation}
and 
\begin{equation}
  \nabla^\dagger \nabla= \nabla\nabla^\dagger,
\end{equation}
we see that (\ref{eqn:interm}) gives 
\begin{equation}
  a \langle \varphi | (\nabla^\dagger \nabla)^\alpha |\varphi \rangle \leq \epsilon
\end{equation}
Reapplying Jensen's inequality, we find that 
\begin{equation}\label{eqn:jen} 
  \langle \varphi | (\nabla^\dagger \nabla)^{\alpha - n} | \varphi \rangle \leq \left(\frac{\epsilon}{a}\right)^{1 - \frac{n}{\alpha}}.
\end{equation}

Once again, we may proceed by calculating
\begin{equation}\label{eqn:latticecom}
  \mathrm{i}[x,H] = \alpha (\nabla^{\dagger})^{\alpha-1} (\nabla^\dagger f + f \nabla) \nabla^{\alpha - 1}
\end{equation}
The form of (\ref{eqn:latticecom}) is different superficially than what one might expect from considering the form of (\ref{eqn:contcom}).
This comes from the fact that the canonical commutation relations are not achievable with any choice of $\nabla$. 
Nonetheless, it is still possible to apply the Cauchy--Schwarz and triangle inequalities in exactly the same manner as before. Explicitly
\begin{equation}
  |\langle \varphi | \mathbb P_\epsilon [x,H]\mathbb P_\epsilon |\varphi \rangle| \leq \alpha |\langle \varphi | \mathbb P_\epsilon (\nabla^\dagger)^\alpha f \nabla^{\alpha - 1} \mathbb P_\epsilon |\varphi \rangle | +  \alpha |\langle \varphi | \mathbb P_\epsilon (\nabla^\dagger)^{\alpha - 1} f \nabla^\alpha \mathbb P_\epsilon | \varphi \rangle | .
\end{equation}
and the Cauchy-Schwarz inequality gives 
\begin{equation}
  |\langle \varphi | \mathbb P_\epsilon (\nabla^\dagger)^\alpha f \nabla^{\alpha-1} \mathbb P _\epsilon |\varphi \rangle | \leq 
  \left(
    \langle \varphi | \mathbb P_\epsilon (\nabla^\dagger)^\alpha f \nabla^\alpha \mathbb P_\epsilon | \varphi \rangle \langle \varphi | \mathbb P_\epsilon (\nabla^\dagger)^{\alpha - 1} f \nabla^{\alpha - 1} \mathbb P _\epsilon | \varphi \rangle
  \right)^{\frac{1}{2}}
\end{equation}
where one factor is bounded above by $\epsilon^{\frac{1}{2}}$ by the definition of $\mathbb P_\epsilon $ and the other factor is bounded above by  $\sqrt{b} \left(\frac{\epsilon}{a}\right)^{\frac{1}{2} - \frac{1}{2 \alpha}}$ due to   (\ref{eqn:jen}).
Continuing as before produces 
\begin{equation}\label{eqn:latticeb}
  \left| \frac{\mathrm d }{\mathrm d t} \langle \varphi | \mathbb P_\epsilon x\mathbb P_\epsilon |\varphi \rangle \right| \leq 2 \alpha \sqrt{b} a^{\frac{1 - \alpha}{2 \alpha}}\epsilon^{1 - \frac{1}{2\alpha}} \langle \varphi | \mathbb P_\epsilon | \varphi \rangle 
\end{equation}
for any state $|\varphi \rangle$. Since $\epsilon \sim T$, this reproduces the scaling of (\ref{eq:vB}).

  We remark that the bound (\ref{eqn:latticeb}) applies to a surprisingly general class of lattice models.
  The simplest such example consists of models  of the form
  \begin{equation}\label{eqn:example}
      H = \sum_{n = -\infty}^\infty c_n |n \rangle \langle n| - |n+1 \rangle \langle n | - |n \rangle \langle n+1| 
  \end{equation}
  satisfy (\ref{eqn:latticeb}) so long as $c_n \geq 2$ for all $n$ in $\mathbb{Z}$. 
  The model (\ref{eqn:example}) appears in the classic single band hopping problem and the bound (\ref{eqn:latticeb}) for this model can be interpreted as a requirement that all dynamics slow down at low energy. 
  If $c_n = 2$ for all $n$, $H$ is diagonalized by plane waves and has dispersion $E(k) = 2(1 - \cos(k))$. Dynamics at low energy are slow because the group velocity 
  $v_G = 2 \sin (k)$, so the result still holds in the presence of a repulsive potential.  The bound (\ref{eqn:latticeb}) generalizes this result. 
\section{Many-body generalization}
The results of Sections \ref{sec:continuum} and \ref{sec:lattice} can be cleanly generalized to the case of $N$ particles in $d$ spatial dimensions, when the Hamiltonian takes a certain form (described shortly). While in this section, we only explicitly treat lattice models in 1d for brevity, the problem in other settings can be handled similarly.

Fix some integer $N > 0$ and for each  $1 \leq i \leq N$ define 
 \begin{equation}
   \nabla_i = \mathbb I ^{\otimes(i-1)} \otimes \nabla\otimes \mathbb I^{\otimes(N - i )}
\end{equation}
and 
\begin{equation}
   x_i = \mathbb I ^{\otimes(i-1)} \otimes x \otimes \mathbb I^{\otimes(N - i )}.
\end{equation}
We may also choose $N$ matrices
\begin{equation}\label{eqn:fmanybod}
  F_i = \sum_{n = -\infty}^{\infty} f_{n,i} \mathbb I^{\otimes(i-1)} \otimes |n \rangle \langle n | \otimes \mathbb I^{\otimes(N-i)}
\end{equation}
with 
\begin{equation}
  0 < a \leq \inf_{n,i} f_{n,i} \leq \sup_{n,i} f_{n,i} \leq b <\infty.
\end{equation}
Naturally, we write 
\begin{equation}\label{eqn:mbham}
  H = \sum_{i = 1}^N (\nabla_i^{\dagger})^\alpha F_i \nabla_i^\alpha + V
\end{equation}
where we demand that for any $i$,
\begin{equation}
    [V,x_i]=0
\end{equation}
and that $V\ge 0$.   

We emphasize this positivity condition on $V$ as crucial.  Intuitively, as in our previous derivation, our derivation of a bound will rely on the inability of the system to gain kinetic energy from attractively-interacting particles getting closer together.

Here, we seek not to bound the velocity of a 
particular particle on the lattice; instead, we wish to constrain the average particle velocity in the worst case at a fixed energy.  This is because, as stated in the introduction, there may be rare states in which energy is highly concentrated in a small number of particles, and any rigorous bound must incorporate that possibility. 
As before, we fix some threshold energy and we restrict to energies only beneath said threshold. However, we wish to consider explicitly extensive 
energies.
Define 
\begin{equation}
  \mathcal E = N \epsilon,
\end{equation}
for $\epsilon \sim N^0$. In all of what follows, we write $\langle \cdot \rangle$ to indicate an expectation value with regard to a particular state, obeying (\ref{eq:Pepsilonprojector}) with $\epsilon$ replaced by total energy $\mathcal{E}$.

Directly following earlier statements, we notice that if \begin{equation}
    a \langle (\nabla_i^\dagger \nabla_i)^\alpha \rangle = \epsilon_i,
\end{equation}
then 
\begin{equation}\label{eqn:lconst}
   \sum_{i=1}^N\epsilon_i   \leq \langle  H \rangle \leq  \mathcal E .
\end{equation}
Proceeding to calculate time derivatives,
\begin{equation}
  [x_i, H] = [x_i , (\nabla^\dagger_i)^\alpha F_i \nabla^\alpha_i ] = 
  \alpha (\nabla_i^\dagger) ^{\alpha - 1}
  (\nabla_i^\dagger F_i + F_i \nabla_i)
  \nabla_i^{\alpha-1}
\end{equation}
and therefore
\begin{equation}
  \left|\frac{\mathrm d}{\mathrm d t}\langle x_i\rangle \right| \leq 2\left(b^2
    \langle (\nabla_i^\dagger \nabla_i)^\alpha\rangle 
    \langle (\nabla_i^\dagger \nabla_i )^{\alpha - 1}\rangle
  \right)^\frac{1}{2},
\end{equation}
while Jensen's inequality gives 
\begin{equation}
  \left|\frac{\mathrm d }{\mathrm d t}\langle x_i \rangle \right| \leq 2 b a^{\frac{1-\alpha}{2\alpha}} \langle (\nabla_i^\dagger \nabla_i)^\alpha \rangle^{1 - \frac{1}{2\alpha}} = 2 b a^{\frac{1-\alpha}{2\alpha}} \left(\frac{\epsilon_i}{a}\right)^{1-\frac{1}{2\alpha}} .
\end{equation}
Now, we seek to maximize the typical velocity \begin{equation}
    v_{\mathrm{typ}} = \frac{1}{N}\sum_{i=1}^N  |\langle v_i\rangle | \le \frac{1}{N}\sum_{i=1}^N 2 b a^{\frac{1-\alpha}{2\alpha}} \left(\frac{\epsilon_i}{a}\right)^{1-\frac{1}{2\alpha}} \label{eq:vtyp}
\end{equation}
over low energy states in which (\ref{eqn:lconst}) holds.  This is readily achieved via Lagrange multipliers, and since the exponent $1-\frac{1}{2\alpha}<1$, we find that the maximal value of the bound on $v_{\mathrm{typ}}$ (\ref{eq:vtyp}) arises when all $\epsilon_i = \mathcal{E}/N$ in (\ref{eq:vtyp}), leading to: \begin{equation}
    v_{\mathrm{typ}} \le b a^{\frac{1}{\alpha} - \frac{3}{2}} \left(\frac{\mathcal E }{N}\right)^{1 - \frac{1}{2\alpha}}. \label{eq:vtypbound}
\end{equation}
Hence, the maximal  average particle velocity in a many body state with total energy $\mathcal E$ scales with average energy per particle in the same manner as a single particle's velocity scales with its energy.  Note that it is important in this result to ask about typical velocity, and not, e.g., root-mean-square velocity\footnote{For any random variable $X$, $\langle X^2 \rangle \geq \langle X\rangle^2$. The typical velocity, $|\langle v\rangle|$ has a finite upper bound in this case, but the same may not be true of the root--mean--squared velocity, $\sqrt{\langle v^2 \rangle }$. In fact, it may be the case that the root--mean--squared velocity is not even finite.}.  Following \cite{Faupin:2021nhk,Faupin:2021qpp}, we deduce that the time it takes to move a macroscopic number of particles a mean displacement of  $L$ cannot be parametrically smaller than $L/v_{\mathrm{typ}}$. 

 This result is largely agnostic to the form of potential chosen, so long as $[V,x_i]=0$, \emph{and} so long as $V\ge 0$ (i.e. the interactions are repulsive).   In particular, one might choose 
 \begin{equation} \label{eq:413}
   V = \sum_{n,m = -\infty}^\infty \sum_{i = 1}^N\sum_{j = i}^N c_{nm,ij} \mathbb I^{\otimes(i - 1)} \otimes |n\rangle \langle n | \otimes \mathbb I^{\otimes(j - i - i)} \otimes |m \rangle \langle m | \otimes \mathbb I^{\otimes(N - j)}
 \end{equation}
 or arbitrary nonlinear combinations of such potentials.  If we make the further choice \begin{equation}
     c_{nm,ij}= U \nabla_{nm},
 \end{equation}
 then the interaction term $V$ becomes the classic Hubbard model \cite{hubbard,Arovas_2022}, albeit so far with distinguishable particles.

 However, notice that particle indistinguishability is actually very straightforward to add in this formalism, so long as $F_i$ is independent of $i$, and the potential $V$ is also permutation symmetric: for example,  $c_{nm,ij} = c_{nm}$ in (\ref{eq:413}). 
 In this setting, one simply restricts to the Bose-Hubbard model or Fermi-Hubbard model by restricting the first-quantized Hilbert space to the fully symmetric or fully antisymmetric irreducible representations of the permutation group.   Therefore we conclude that (\ref{eq:vtypbound}) applies to both the Bose-Hubbard and Fermi-Hubbard models, with either attractive or repulsive interactions.  
 
 In the physically relevant case of a Fermi-Hubbard model with attractive interactions, one expects a Fermi surface to form at low temperature, meaning that the average energy density is finite even at arbitrarily low temperature: \begin{equation}
     \epsilon = \epsilon_0 + cT+ \cdots.
 \end{equation}
 In this case, the typical velocity stays finite even close to the ground state.  This is not surprising -- as noted in the introduction, a theory with a Fermi surface is not necessarily expected to have a vanishing notion of butterfly or typical velocity at low temperatures -- e.g. this does not happen in ordinary Fermi liquid theory.



 Lastly, we remark that a similar bound to (\ref{eq:413}), albeit with a larger prefactor, will hold if we restrict our attention to arbitrary subsets of the $N$ particles as well.  Of course, the prefactor will become exceedingly lousy, as if we only consider a single particle, (\ref{eqn:lconst}) does not forbid that single particle from carrying all of the energy in the system.

\section{Outlook}
In this paper, we described rigorous results on typical particle velocities ``$v_\mathrm B$" in low energy states of lattice and/or continuum models.  Our results generalized to many-particle dynamics in which the kinetic energy takes a single-particle form, and the key result followed from understanding the behavior of low-temperature dynamics in single-particle models.  This is qualitatively similar to the way that the dynamics of bosons has been bounded in \cite{Faupin:2021nhk,Faupin:2021qpp}.

While our bounds hold for \emph{all} low energy states, in the many-body case, we were forced to study only typical particle velocities.  
 We conjecture that our bounds hold for \emph{single particle} velocities if one instead \emph{averages} over low energy states.  More generally, we did not find Lieb-Robinson-like bounds on many-body dynamics.  One reason this is likely to be challenging is that in interacting boson problems (with conserved charge), it is known that signaling and information can propagate parametrically faster than a typical particle \cite{lucas_finite_density,Faupin:2021nhk,Faupin:2021qpp}. Proving a more Lieb-Robinson-like bound on commutator norms may well require qualitatively new technical methods.  Discovering such methods may be the key to generalize the vast literature on Lieb-Robinson bounds \cite{chen2023speed} to derive bounds on low temperature physics.

 Lastly, let us remark briefly on our first attempt to tackle the problem of deriving bounds on dynamics at low energies/temperature.  Intuitively, one might expect that the correct way to bound $v_{\mathrm{B}} \le  T\xi $ is to argue that low-energy dynamics is efficiently generated by some Hamiltonian in which terms have bounded magnitude $T$, at the price of spatial non-locality on the length scale $\xi$.  This would lead to an elegant picture in which $\xi \sim T^{-1/z}$ and $v_{\mathrm{B}}\sim T\xi$.  Unfortunately we were not able to find a way to realize this intuition rigorously, even in the single-particle setting.  Along these lines, the Appendix presents a more physically intuitive derivation of known results on the locality of the density matrix for single-particle Hamiltonians.  It would be interesting if future work can present an alternative derivation of the bounds of this paper from this perspective.

\section*{Acknowledgements}
This work was supported by a research Fellowship from the Alfred P. Sloan Foundation under Grant FG-2020-13795, and by the U.S. Air Force Office of Scientific Research under Grant FA9550-21-1-0195.

\begin{appendix}
\section{Locality of the canonical density matrix for a single particle}
In this appendix, we use mathematical techniques inspired by our recent ``quantum walk bound" method \cite{chen2023speed,Lucas:2019cxr} to derive bounds on the locality of the (unnormalized) matrix elements of the density matrix $\mathrm{e}^{-\beta H}$, for single-particle problems at large $\beta$.     We emphasize to the reader that these bounds do not \emph{directly} bound the elements of the thermal \emph{density matrix}, which must be normalized by the single-particle partition function.

First, we will consider only single particle models of the form 
\begin{equation}
  H = \nabla^{\dagger \alpha } \nabla^\alpha + V
\end{equation}
with  $V$ positive semidefinite.
We wish to show, for a positive number $\beta$, 
 \begin{equation}
   |\langle n | \mathrm{e}^{-\beta H} |m \rangle | \leq c \mathrm{e}^{-\frac{|n-m|}{\xi}}
\end{equation}
for a particular $\beta$--dependent correlation length  $\xi$ and constant $c$.  Notice that if we divide this matrix element by the partition function, this inequality tells us that thermal single-particle correlation functions decay exponentially with distance.

Define 
 \begin{equation}
   F = \sum_{n = -\infty}^\infty \mathrm{e}^{\mu n} |n \rangle \langle n |. 
\end{equation}
We require 
\begin{equation}
  [F,V ] = 0 
\end{equation}
which would also be guaranteed by  $V = V(x)$. 
Explicitly, we seek to show
\begin{equation}\label{eqn:walkode}
  \frac{\mathrm d }{\mathrm d \beta }\langle \varphi | \mathrm{e}^{-\beta H} F \mathrm{e}^{-\beta H} |\varphi \rangle  \leq h(\mu) \langle \varphi | \mathrm{e}^{-\beta H } F \mathrm{e}^{-\beta H } |\varphi \rangle 
\end{equation}
in order to bound the evolution in $\beta$ of a particular  element of $\mathrm{e}^{-\beta H}$ by solving the differential equation (\ref{eqn:walkode}).

Define 
\begin{equation}
  |\phi \rangle = F^{\frac{1}{2}} \mathrm{e}^{-\beta H } |\varphi \rangle.
\end{equation}
and observe that 
\begin{equation}
  \frac{\mathrm d}{\mathrm d \beta } \langle \phi | \phi \rangle = - \langle \phi | F^{-\frac{1}{2}}\{F,H\}F^{-\frac{1}{2}} |\phi \rangle = - \langle \phi | F^{-\frac{1}{2}} \{F, H - V\} F^{-\frac{1}{2}} |\phi \rangle - \langle \phi | V|\phi \rangle .
\end{equation}
$V$ is taken to be positive definite, so we find that 
\begin{equation}
  \frac{\mathrm d }{\mathrm d \beta } \langle \phi | \phi \rangle \leq  -\inf_{|\lambda\rangle} \langle \lambda|F^{-\frac{1}{2}} \{F,H - V\} F^{-\frac{1}{2}} |\lambda \rangle \langle \phi | \phi \rangle.
\end{equation}
Since $F^{-\frac{1}{2}}\{H-V,F\} F^{-\frac{1}{2}}$ is translation invariant, it is exactly diagonalized by plane waves. We find 
\begin{equation}\label{startarg}
  \frac{\mathrm d }{\mathrm d \beta } \langle \phi |\phi \rangle \leq 4^\alpha \sinh^{2\alpha}\left(\frac{\mu}{2}\right) \langle \phi |\phi \rangle.
\end{equation}
Now, we simply choose $|\varphi \rangle = |n \rangle$, and we find 
 \begin{equation}
   |\langle n | \mathrm{e}^{-\beta H} | m\rangle |^2 \leq \mathrm{e}^{4^\alpha \sinh^{2\alpha}(\mu/2)\beta - \mu |n - m| }.
\end{equation}
If we then choose $\mu = \frac{1}{\beta^{\frac{1}{2\alpha}}}$, and 
use $\sinh(\mu/2)\le 2^{1/(2\alpha)} \mu/2$ at sufficiently small $\mu$ (which corresponds to large $\beta$, i.e. small $T$) we see that 
 \begin{equation}
   |\langle n | \mathrm{e}^{-\beta H }  | m \rangle| \leq  \mathrm{e}^{ 2- \frac{|n-m|}{\xi}}
\end{equation}
with 
\begin{equation}\label{endarg}
  \xi = \beta^{\frac{1}{2\alpha}}.
\end{equation}

This ``quantum walk bound" also shows that a generic positive semidefinite, $d$--local, bounded Hamiltonian $H$ has a bounded correlation length.
We need only recognize 
\begin{equation}
    -\frac{\mathrm d }{\mathrm d \beta} \langle \phi | \phi\rangle = \langle \phi| F^{-\frac{1}{2}} H F^{\frac{1}{2}} + F^{\frac{1}{2}} H F^{-\frac{1}{2}}|\phi\rangle =2 \langle \phi | H |\phi\rangle + \sum_{n,m} 2(\cosh\left(\frac{\mu (n-m)}{2}\right)-1) \langle \phi |n \rangle \langle n | H |m\rangle \langle m |\phi\rangle.
\end{equation}
As before, we seek to bound $\frac{\mathrm d }{\mathrm d \beta}   \langle \phi|\phi\rangle $ from above. By assumption $\langle \phi | H |\phi\rangle$ is nonnegative so 
\begin{align}
    \frac{\mathrm d }{\mathrm d \beta } \langle \phi | \phi \rangle &\leq -2 
    \sum_{n,m} 2(\cosh\left(\frac{\mu (n-m)}{2}\right)-1) \langle \phi |n \rangle \langle n | H |m\rangle \langle m |\phi\rangle \notag \\
    &\leq 
    \sum_{n} |\langle \phi|n\rangle |^2 \sum_{m} \left|
    \cosh\left( \frac{\mu(n-m)}{2}\right) -  1
    \right| |\langle n | H | m \rangle |. 
\end{align}
Since $H $ has bounded-range interactions, there must exist some $J > 0$ so that 
\begin{equation}
    \sup_n\sum_m |\cosh(\frac{\mu}{2}(n-m)) -1| |\langle n | H | m \rangle | \leq \mu^2 J 
\end{equation}
for sufficiently small $\mu$. In particular, 
\begin{equation}
    \sup_n \sum_{m} |n-m|^2 |\langle n | H | m \rangle | \leq 2 d \sup_{n,m} |\langle n | H | m \rangle |. 
\end{equation}
The reasoning of (\ref{startarg})-(\ref{endarg})  can be repeated to show that there exists some $c > 0$ so 
\begin{equation}\label{eq:xi=sqrtbeta}
    \xi \leq  c \sqrt{\beta}
\end{equation}
in general. 
The exponential decay of correlations in a thermal state is a well--known result achieved previously through e.g. the use of Chebyshev polynomial asymptotics \cite{rho_sqrt97,saito_finite_temp}: see Theorem 18 in the recent \cite{QSVT23}, where to approximate the exponential function with small error in the so called regime 2, it is sufficient to truncate the Chebyshev series at order \eqref{eq:xi=sqrtbeta}. This technical method may be employed even in many-body systems.  The above derivation focuses on the single-particle problem, but is perhaps more physically transparent.

\end{appendix}

\bibliography{thebib}

\end{document}